\documentclass[aps,prd,notitlepage,showpacs]{revtex4-2}

\usepackage{graphicx}
\usepackage{amssymb}
\usepackage{amsmath}
\usepackage{hyperref}
\usepackage{bm}
\usepackage{latexsym}
\usepackage{epsfig}
\usepackage{psfrag}
\usepackage[normalem]{ulem}
\usepackage{textcomp}
\usepackage{color}
\usepackage{pstricks}
\usepackage[utf8]{inputenc}
\usepackage{natbib}

\makeatletter
\gdef\@fpheader{}
\makeatother

\def\nn{\nonumber} 
\def\pa{{\partial}}
\def\f{\frac}
\def\l{\left}
\def\r{\right}

\def\cA{\mathcal A}

\def\Mp{M_{_{\rm Pl}}}

\def\cR{{\mathcal R}}
\newcommand{\g}{\gamma}
\newcommand{\viz}{\textit{viz.~}}

\begin{document}
\title{A gauge invariant prescription to avoid $\g$-crossing instability in Galileon bounce}
\author{Rathul Nath Raveendran}\email{rathulnath.r@gmail.com}
\affiliation{The Institute of Mathematical Sciences, CIT Campus,
Chennai~600113, India}
\begin{abstract}
We revisit the evolutions of scalar perturbations in a
non-singular Galileon bounce.
It is known that the second order differential equation governing the perturbations
is numerically unstable at a point called $\g$-crossing.
This instability is usually circumvented using certain gauge choices.
We show that the perturbations can be evolved across this point by solving the first order differential equations governing suitable gauge invariant quantities without any instabilities. We demonstrate this method in a matter bounce scenario described by
the Galileon action.
\end{abstract}
\maketitle

\section{Introduction}
The most investigated alternative to the inflationary paradigm is the classical
non-singular bouncing scenario. In a non-singular bouncing scenario, the universe undergoes a period of contraction
until the scale factor attains a minimum value and thereafter it transits to the phase of
expansion.
The Hubble parameter is negative during the contraction, and it
is positive during the expansion.
These two phases are connected by a bouncing phase where the rate of change
of the Hubble parameter is positive.
In Einsteinian gravity this phase is obtained by violating the null energy condition (NEC).
During this NEC violating phase, the Hubble parameter grows and reaches zero and continues to grow until it reaches a positive value and then it begins to decrease.
It is well known that a canonical scalar field with quadratic kinetic term does not violate the NEC.
To achieve this condition, various models have been considered~\cite{Brandenberger:2016vhg}. One of the best motivated
model is the cubic Galileon bounce~\cite{Qiu:2011cy,Easson:2011zy,Brandenberger:2016vhg,Akama:2017jsa,Cai:2017tku}.
\par
It has been recognized that in the Galileon bounce models, there is a crossing point
\viz $\g$-crossing,
where the equation of motion governing the perturbation becomes numerically unstable~\cite{Battarra:2014tga,Ijjas:2017pei,Ijjas:2016tpn}.
It is also understood that this is not a real divergence in the perturbation.
Usually, this instability can be evaded by choosing a particular gauge such as
harmonic gauge or Newtonian gauge to evolve the perturbations~\cite{Battarra:2014tga,Ijjas:2017pei,Mironov:2018oec}.
In this work, we demonstrate that the primordial perturbations can be evolved across the $\g$-crossing by solving two first order differential equations governing
suitable gauge invariant quantities.
\par
We shall work with natural units such that $\hbar=c=1$, and set the Planck mass 
to be $\Mp=\l(8\,\pi\, G\r)^{-1/2}$. 
We shall adopt the metric signature of $\l(-, +, +, +\r)$. 
Note that, while Greek indices shall denote the spacetime coordinates, the Latin 
indices shall represent the spatial coordinates, except for $k$ which shall be 
reserved for denoting the wavenumber. 
Moreover, an overdot and an overprime shall denote differentiation with respect 
to the cosmic and the conformal time coordinates, respectively.
\par
This paper is organized as follows. In the following section, we shall briefly discuss
the Galileon bounce.
In section~\ref{sec-2}, we shall obtain the relevant first order equations
governing the perturbations. In section~\ref{sec-3}, we shall use these equations to evolve the scalar perturbations in a particular bouncing scenario called
matter bounce. We shall conclude in section~\ref{sec-4} with brief summary and
outlook.
\section{The model}\label{sec-1}
We assume that the bounce stage is driven by a single scalar field $\phi$ that is described by the generalized cubic Galileon action~\cite{Brandenberger:2016vhg,Qiu:2011cy,Easson:2011zy,Ijjas:2016tpn}
\begin{equation}\label{eq:S}
S=\int{\rm d}^4 \,x\,\sqrt{-g}\, \cal{L},
\end{equation}
with the Lagrangian density
\begin{equation}
\label{eq:G-Lagrangian}
{\cal L}=K\left(X,\phi \right) -b(\phi) X \Box\phi,
\end{equation} 
where
\begin{equation}
X =-\f{1}{2}\,\pa_{\mu} \phi \,\pa^{\mu} \phi,
\end{equation}
and $b(\phi)$ is the dimensionless coupling of the scalar field $\phi$ to the Galileon term. 
We shall consider the background to be the spatially flat,
Friedmann-Lema\^itre-Robertson-Walker (FLRW) metric that is described by the line-element
\begin{equation}\label{eq:metric-bg}
{\rm d} s^2=-{\rm d}t^2+a^2(t)\,\delta_{ij} \, {\rm d}x^i\,{\rm d}x^j,
\end{equation}
where $a(t)$ is the scale factor. Varying the action (\ref{eq:S}) with respect to the background metric we get the Friedmann equations in terms of the background quantities as
\begin{subequations}
\begin{eqnarray}
\label{eq:H21}
3 \, \Mp^2 \, H^2  &=&- K\,+ K_X\, \dot{\phi}^2
- \frac{1}{2}\,b_{\phi}\,\dot{\phi}^4
+3\, H\, b\,\dot{\phi}^3,\quad \\
\label{eq:Hd1}
-2\Mp^2 \dot{H} &=& K_X \, \dot{\phi}^2  -b_{\phi}\, \dot{\phi}^4
 + 3\,H\,b \, \dot{\phi}^3 -  b\,\ddot{\phi}\,\dot{\phi}^2   
,\quad
\end{eqnarray}
\end{subequations}
where subscripts $\phi$ and $X$ denote the differentiation with respect to $\phi$ and $X$ respectively. By choosing the functions $K(\phi,\chi)$ and $b(\phi)$ suitably, one
can model the early contracting phase, the NEC violating phase and the expanding phase of the universe~\cite{Brandenberger:2016vhg,Qiu:2011cy,Easson:2011zy,Ijjas:2016tpn}.
\section{Evolution of perturbations}\label{sec-2}
If we take into account the scalar perturbations to the background 
metric~(\ref{eq:metric-bg}), then the FLRW line-element, in general,  
can be written as~\cite{Mukhanov:1990me}
\begin{equation}
{\rm d} s^2
= -\left(1+2\, A\right)\,{\rm d} t ^2 
+ 2\, a(t)\, (\partial_{i} B)\; {\rm d} t\; {\rm d} x^i\,+a^{2}(t)\; \left[(1-2\, \psi)\; \delta _{ij}
+ 2\, \left(\partial_{i}\, \partial_{j}E \right)\right]\,
{\rm d} x^i\, {\rm d} x^j, \label{eq:f-le-sp}
\end{equation}
where $A$, $B$, $\psi$ and $E$ are four scalar functions that describe 
the perturbations, which depend on time as well as space.
\par
In order to identify the suitable gauge invariant quantities to evolve the perturbations,
it is important to understand how each of the basic perturbations, $A$, $B$, $\psi$, $E$ and $\delta \phi$, transform under the
coordinate transformations.
It is easy to show that, under the coordinate transformations
\begin{equation}\label{eq:c-t}
t \to t+ \delta t,\quad x_i \to x_i + \pa_i \delta x,
\end{equation}
the functions $A$, $B$, $\psi$ 
and $E$ transform as follows:
\begin{subequations}
\begin{eqnarray}
A &&\to A - \dot{\delta t},\\
B &&\to B + \l(\delta t/a\r) - a\, \dot{\delta x},\\
\psi && \to  \psi + H\, \delta t,\label{eq:psit}\\
E&&\to E-\delta x.\label{eq:sgt-mv}
\end{eqnarray}
\end{subequations}
It is evident from the expression~(\ref{eq:psit}) that the spatially flat gauge, wherein $\psi=0$, is ill-defined at the bounce where $H=0$.
It is convenient to define $\sigma=a\,B-a^2\,\dot{E}$ and this quantity transforms as
\begin{equation}
\sigma \to \sigma+ \delta t.
\end{equation}
The perturbation in the scalar field transforms as
\begin{equation}
\delta \phi \to \delta \phi-\dot{\phi} \, \delta t.
\end{equation}
Using the above information, we can define the following three gauge invariant quantities as
\begin{subequations}
\begin{eqnarray}
\cR &=& \psi + H \,\f{\delta \phi}{\dot{\phi}},\label{eq:R-def}\\
\Sigma &=& \sigma + \f{\delta \phi}{\dot{\phi}},\label{eq:Sigma-def}\\
\cA &=& A - \l(\f{\delta \phi}{\dot{\phi}}\r)^{\cdot}.\label{eq:A-def}
\end{eqnarray}  
\end{subequations}
At the first order in the perturbations, three of the Einstein's equations
of our interest are given by~\cite{Mukhanov:1990me,Sriramkumar:2009kg}
\begin{subequations}\label{eq:fo-ee}
\begin{eqnarray}
	3\,H\,\l(H\,A + \dot{\psi}\r) 
	- \frac{
	\nabla^2}{a^2}\l[\psi - H\,\sigma\r]
	&=& -\frac{\delta T^0_0}{2\,\Mp^2},\\
	\pa_i\l(H\,A + \dot{\psi}\r) 
	&=& -\frac{\delta T^0_i}{2\,\Mp^2},\\
	A - \psi + \frac{1}{a}\,\l(a\,\sigma\r)^{\cdot}&=&0\label{eq:fo-ee-3},
\end{eqnarray}
\end{subequations}
where $\delta T^\mu_\nu$ is the perturbed stress-energy tensor associated with the Galileon.
The equation (\ref{eq:fo-ee-3}) follows from the fact that there are no anisotropic
stresses present.
The components of $\delta T^\mu_\nu$ can be calculated from the action
(\ref{eq:S}) to be
\begin{subequations}
\begin{eqnarray}\label{eq:delta-T}
\frac{\delta T^0_0}{2\,\Mp^2} &=& -\bigg[\l(\theta+3\,H^2
-3\,\g^2\r)A+\f{1}{2}\,b\,\dot{\phi}^3\l( 3\, \dot{\psi}
+\f{\nabla^2}{a^2}\,\sigma \r)\nn \\ 
& &- \l(\theta+3\,H\,\g-3\,\g^2\r)\l(\f{\delta \phi}{\dot{\phi}}\r)^\cdot
-\l(3\,\dot{H}\,\g-\f{\nabla^2}{2\,a^2}\, b\,\dot{\phi}^3\r) \f{\delta \phi}{\dot{\phi}}\bigg],\\
\frac{\delta T^0_i}{2\,\Mp^2}&=& -\partial_i \l[ \f{1}{2}\,b\,\dot{\phi}^3\, A-
\l(H-\g\r)\l(\f{\delta \phi}{\dot{\phi}}\r)^\cdot
-\dot{H}\,\f{\delta \phi}{\dot{\phi}}\r],
\end{eqnarray}
\end{subequations}
where
\begin{subequations}\label{eq:gamma-theta}
\begin{eqnarray}
\g &=& H - \f{1}{2}\,b \, \dot{\phi}^3 \label{eq:gamma},\\
\theta &=& 3 \l(\g-H\r)^2-\dot{H}+\f{1}{2}\,b\, \ddot{\phi}\, \dot{\phi}^2
+\f{1}{2}\, K_{XX} \dot{\phi}^4 + \f{3}{2}\,b\,H\,\dot{\phi}^3 - \f{1}{2}\,b_{\phi} \dot{\phi}^4 \label{eq:sigma}.   
\end{eqnarray}
\end{subequations}
We define $\delta T^0_i=-\partial_i\delta q$ and the quantity $\delta q$ transforms under coordinate transformations (\ref{eq:c-t}) as
\begin{equation}
\delta q \to \delta q+2\,\dot{H} \delta t. 
\end{equation}
It is important to note that the comoving gauge, where $\delta q=0$, is ill-defined at $
\dot{H}=0$.

\par

The scalar part of the linearized Einstein equations~(\ref{eq:fo-ee}) can be rewritten in terms of the gauge invariant quantities defined in equations~(\ref{eq:R-def}),
(\ref{eq:Sigma-def}) and (\ref{eq:A-def}) as
\begin{subequations}\label{eq:Es-r-a-sigma}
\begin{eqnarray}
\cR' + \f{k^2\, \g}{a \, \theta} \, \l( \cR - \g \, \Sigma\r)&=&0,\label{eq:Es-r}\\
\cR' + a\, \g\, \cA &=&0,\label{eq:Es-a}\\
\Sigma' +a\,  H\, \Sigma + \f{k^2}{a \, \theta} \, \l( \cR - \g \, \Sigma \r)-a \, \cR&=&0.\label{eq:Es-sigma}
\end{eqnarray}
\end{subequations}
Note that the above equations are written in terms of conformal time coordinate.
From equations~(\ref{eq:Es-r}) and (\ref{eq:Es-a}), it is evident that $\cR'=0$ at $\g=0$. One can easily obtain the second order differential equation governing $\cR$ from equations~(\ref{eq:Es-r})
and (\ref{eq:Es-sigma}) as
\begin{equation}\label{eq:R''}
\cR''+2\,\f{z'}{z}\, \cR' + k^2\,c_s^2 \, \cR=0, 
\end{equation} 
where,
\begin{eqnarray}
z^2 &=& 2 \, a^2 \f{\theta}{\g^2},\\
c_s^2 &=& \f{\g^2}{\theta}\,\l(\f{H}{\g}-\f{\g'}{a\,\g^2}-1\r).
\end{eqnarray}
It should be mentioned that the quantity $\theta$ requires to be positive to avoid the
quantum ghost instabilities~\cite{Lehners:2008vx,Ijjas:2016tpn}.
It is evident from the above expressions that, the coefficient of $\cR'$ 
\begin{equation}
2\, \f{z'}{z}=2\, a\, H+\f{\theta'}{\theta}-2\,\f{\g'}{\g},
\end{equation}
diverges at $\g=0$.
This is not a concern, since, as we have argued earlier, $\cR'$ becomes zero at this point.
In other words, from equations~(\ref{eq:Es-r}) and (\ref{eq:Es-a}), it is evident that the ratio $\cR'/\gamma$ is finite.
This implies that there is no divergence in $\cR$ at $\g$-crossing.
However, the equation ($\ref{eq:R''}$) can be numerically unstable around $\g=0$ (for a discussion on this point, see Refs.~\cite{Battarra:2014tga,Ijjas:2017pei,Ijjas:2016tpn}).
In order to avoid this stiffness in the equation one can use equations (\ref{eq:Es-r}) and (\ref{eq:Es-sigma}) for evolving the perturbations. In the following section, we shall illustrate this point with a specific example.


\section{Matter bounce}\label{sec-3}
In order to show that the quantity $\cR$ evolves smoothly across the point where $\g=0$ without any divergence, we solve the equations (\ref{eq:Es-r}) and (\ref{eq:Es-sigma}) for a bounce model called the matter bounce.
Matter bounce is a popular alternative to inflation due to the fact that it generates
scale invariant perturbations as in the case of inflation~\cite{Battefeld:2014uga,Brandenberger:2012zb}.
In matter bounce scenarios, during
the early stages of the contracting phase, the scale factor behaves as in a matter dominated universe. 
In this work, we choose to work with a scale factor of the form~\cite{Raveendran:2017vfx}
\begin{equation}
a(\eta) = a_0\l(1 + k_0^2\,\eta^2\r),
\label{eq:sf}
\end{equation}
where $a_0$ is the value of the scale factor at the bounce and $k_0$ is the scale associated with the bounce.
The above scale factor describes the evolution of matter dominated contracting phase, the bounce phase and the expanding phase up to the beginning of radiation dominated epoch.
Using this scale factor the Hubble parameter and its time derivative can be calculated in terms of scale factor as
\begin{subequations}
\begin{eqnarray}
H^2= \l(\f{2 k_0 }{a_0}\r)^2 \l[\f{1}{(a/a_0)^3}-\f{1}{(a/a_0)^4}\r],\\
\label{eq:H2a}
\dot{H}= -\l(\f{k_0}{a_0}\r)^2 \l[\f{6}{(a/a_0)^3}-\f{8}{(a/a_0)^4}\r].
\label{eq:Hda}
\end{eqnarray}
\end{subequations}
\par
We consider the conventional form of Galileon action with
\begin{equation}
K(X,\phi)=\alpha(\phi)\,X + \beta(\phi)\,X^2-V(\phi),
\end{equation}
where $\alpha({\phi})$ and $\beta({\phi})$ are functions of $\phi$ and $V(\phi)$ is the scalar potential~\cite{Easson:2011zy}.
For simplicity, we assume
\begin{equation}\label{eq:phid2}
\dot{\phi}^2=\l(\f{c_0 k_0}{a_0}\r)^2 \f{1}{(a/a_0)^3},
\end{equation}
where $c_0$ is a constant.
We also set $V(\phi)=0$ and $b$ to be a constant as $b=b_0/(c_0^3\,a_0^2\,\,k_0^2)$.
Upon using equations (\ref{eq:gamma-theta}), (\ref{eq:sf}) and (\ref{eq:phid2}), one can  obtain
\begin{subequations}\label{eq:gamma-theta-mb}
\begin{eqnarray}
\g &=& \f{a_0\, k_0}{a^2}\,\l( 2\, k_0 \, \eta - \f{b_0}{2\, a^{5/2}\,a_0^{3/2}}\r) ,\label{eq:gamma-mb}\\
\theta&=& \f{3\,a_0^2\,k_0^2}{4\,a^4}\bigg[\l(3\, k_0\, \eta - \f{b_0}{a^{5/2}\, a_0^{3/2}} \r)^2
+8+ 31\,k_0^2\,\eta^2\bigg].\label{eq:theta-mb}
\end{eqnarray}
\end{subequations}
\begin{figure}[h!]
	\begin{center}
		\includegraphics[width=8.5cm]{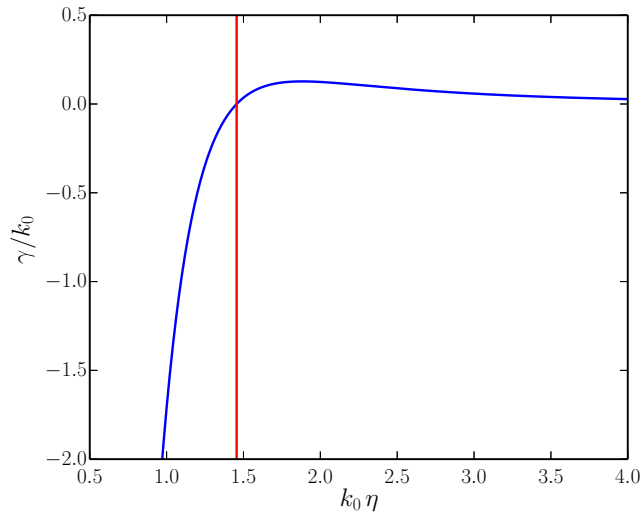}
		\caption{\label{Fig:gamma}Evolution of $\g/k_0$ for $b_0=100$, $a_0=1$ near $\g$-crossing.
			The red vertical line denotes the time when $\g=0$ ($\g$-crossing).}
	\end{center}
\end{figure}
The evolution of $\g$ around the $\g$-crossing in our model of interest is shown in the figure~\ref{Fig:gamma}. It is clear from the expression (\ref{eq:theta-mb}) that in our model, $\theta$ is always positive and hence there is no appearance of ghost instability.
\par
Let us now integrate the equations (\ref{eq:Es-r}) and (\ref{eq:Es-sigma}) numerically
by using the analytical expressions for the background quantities $\g$ and $\theta$.
The initial conditions of $\cR$ can be obtained from the Bunch-Davies initial
condition as  
\begin{equation}
\cR_i=\f{1}{z\,\sqrt{2 \,c_s \,k}} \, {\rm e}^{- i \, c_s\,k \,\eta}\label{eq:Ri}.
\end{equation}
Similarly the initial condition for the quantity $\Sigma$ can be obtained from the
relation~(\ref{eq:Es-r}).
\begin{figure}
	\begin{center}
		\includegraphics[width=.48\textwidth]{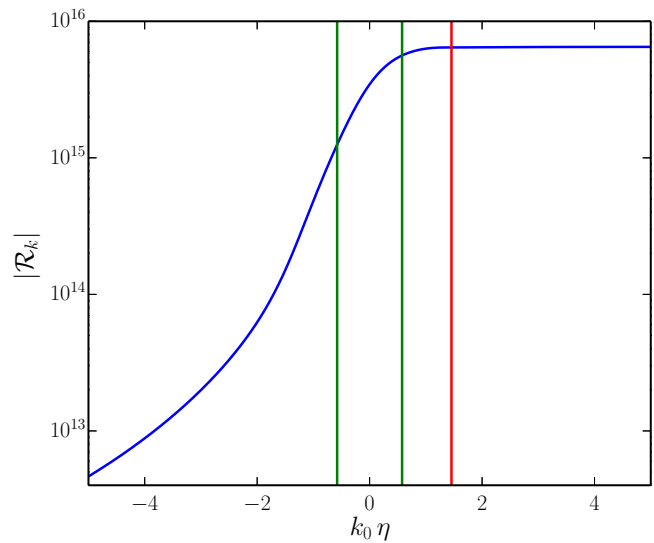}
		\caption{\label{Fig:R-num}Evolution of $\cR_k$ for $k=10^{-10}\,k_0$, $b_0=100$, $a_0=1$ near $\g$-crossing.
		Two green vertical lines drawn before and after the bounce denote
		the beginning and the ending of the NEC violating phase.
		The red vertical line denotes the time when $\g=0$.
		Clearly, there is no divergence in $\cR$ at
	    $\g=0$.}
	\end{center}
\end{figure}
\par
\begin{figure}
	\begin{center}
		\includegraphics[width=.48\textwidth]{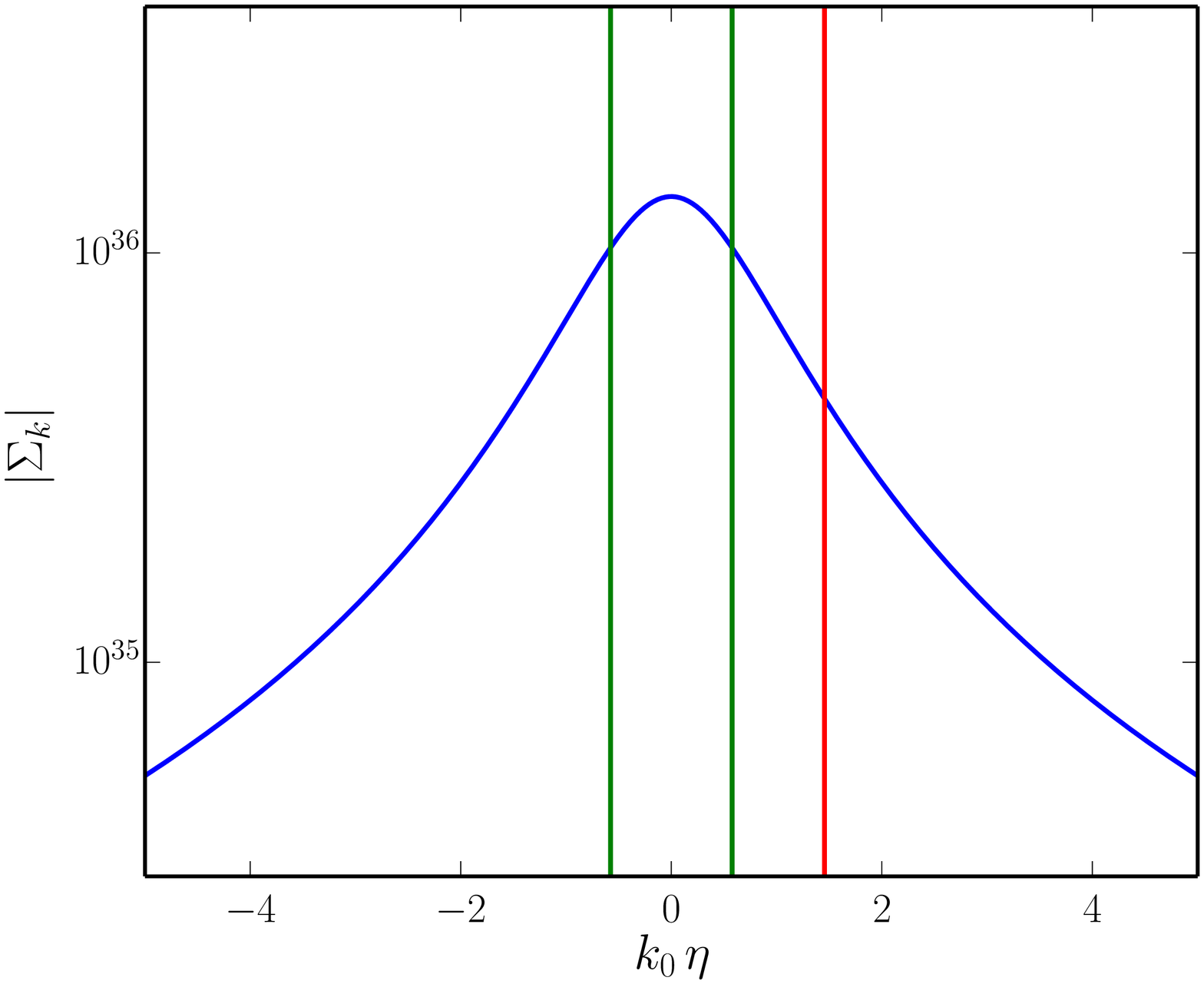}
		\caption{\label{Fig:Sigma-num}Evolution of $\Sigma_k$ for $k=10^{-10}\,k_0$, $b_0=100$, $a_0=1$ near $\g$-crossing.
			Two green vertical lines before and after the bounce denote
			the beginning and ending of the NEC violating phase. The red vertical line denotes the time when $\g=0$.}
	\end{center}
\end{figure}

The figure~\ref{Fig:R-num} shows the evolution of the quantity $\cR$ which is
obtained by integrating the equations (\ref{eq:Es-r}) and (\ref{eq:Es-sigma}).
It is clear from the figure that the quantity $\cR$ evolves smoothly across the point where $\g=0$ ($\g$-crossing).
The evolution of $\Sigma$ is shown in figure~\ref{Fig:Sigma-num}.
\section{Discussion}\label{sec-4}
In this work we have found that the commonly used gauges,
such as spatially flat and comoving gauges,
cannot be used for studying the evolution of the perturbations in bouncing scenarios.
More specifically, while the spatially flat gauge is ill-defined at the bounce, the comoving gauge
is ill-defined at $\dot{H}=0$.
\par
For studying the evolution of perturbations in the Galileon bounce, the usual practice is to use a gauge
in which the equation governing the perturbation does not show any divergence at $\g$-crossing.
In this work, we have demonstrated that the gauge invariant perturbation $\cR$, defined in the equation~(\ref{eq:R-def}), can be used to evolve the perturbation
in Galileon bounce. Moreover, we have also shown that, in order to avoid the numerical instability
near the $\g$-crossing, one can solve the first order equations~(\ref{eq:Es-r-a-sigma}) for studying the evolution of $\cR$.
We have explicitly illustrated this point in a matter bounce model
which is described by the scale factor of the form~(\ref{eq:sf}).
\par
We believe that we can extend this method of evolving the perturbations
across the $\g$-crossing
using the first order equations in models derived from more general Horndesky theories~\cite{Ijjas:2017pei}. We are presently investigating this possibility.
\vskip 6pt\noindent
\underline{\it Acknowledgements:}\/
The author would like to thank L.~Sriramkumar, Ghanashyam Date and Krishnamohan Parattu for comments on the manuscript.
\bibliography{mb-G-2019}

\begin{thebibliography}{15}%
\makeatletter
\providecommand \@ifxundefined [1]{%
 \@ifx{#1\undefined}
}%
\providecommand \@ifnum [1]{%
 \ifnum #1\expandafter \@firstoftwo
 \else \expandafter \@secondoftwo
 \fi
}%
\providecommand \@ifx [1]{%
 \ifx #1\expandafter \@firstoftwo
 \else \expandafter \@secondoftwo
 \fi
}%
\providecommand \natexlab [1]{#1}%
\providecommand \enquote  [1]{``#1''}%
\providecommand \bibnamefont  [1]{#1}%
\providecommand \bibfnamefont [1]{#1}%
\providecommand \citenamefont [1]{#1}%
\providecommand \href@noop [0]{\@secondoftwo}%
\providecommand \href [0]{\begingroup \@sanitize@url \@href}%
\providecommand \@href[1]{\@@startlink{#1}\@@href}%
\providecommand \@@href[1]{\endgroup#1\@@endlink}%
\providecommand \@sanitize@url [0]{\catcode `\\12\catcode `\$12\catcode
  `\&12\catcode `\#12\catcode `\^12\catcode `\_12\catcode `\%12\relax}%
\providecommand \@@startlink[1]{}%
\providecommand \@@endlink[0]{}%
\providecommand \url  [0]{\begingroup\@sanitize@url \@url }%
\providecommand \@url [1]{\endgroup\@href {#1}{\urlprefix }}%
\providecommand \urlprefix  [0]{URL }%
\providecommand \Eprint [0]{\href }%
\providecommand \doibase [0]{https://doi.org/}%
\providecommand \selectlanguage [0]{\@gobble}%
\providecommand \bibinfo  [0]{\@secondoftwo}%
\providecommand \bibfield  [0]{\@secondoftwo}%
\providecommand \translation [1]{[#1]}%
\providecommand \BibitemOpen [0]{}%
\providecommand \bibitemStop [0]{}%
\providecommand \bibitemNoStop [0]{.\EOS\space}%
\providecommand \EOS [0]{\spacefactor3000\relax}%
\providecommand \BibitemShut  [1]{\csname bibitem#1\endcsname}%
\let\auto@bib@innerbib\@empty
\bibitem [{\citenamefont {Brandenberger}\ and\ \citenamefont
  {Peter}(2016)}]{Brandenberger:2016vhg}%
  \BibitemOpen
  \bibfield  {author} {\bibinfo {author} {\bibfnamefont {R.}~\bibnamefont
  {Brandenberger}}\ and\ \bibinfo {author} {\bibfnamefont {P.}~\bibnamefont
  {Peter}},\ }\bibfield  {title} {\bibinfo {title} {{Bouncing Cosmologies:
  Progress and Problems}},\ }\href@noop {} {\  (\bibinfo {year} {2016})},\
  \Eprint {https://arxiv.org/abs/1603.05834} {arXiv:1603.05834 [hep-th]}
  \BibitemShut {NoStop}%
\bibitem [{\citenamefont {Qiu}\ \emph {et~al.}(2011)\citenamefont {Qiu},
  \citenamefont {Evslin}, \citenamefont {Cai}, \citenamefont {Li},\ and\
  \citenamefont {Zhang}}]{Qiu:2011cy}%
  \BibitemOpen
  \bibfield  {author} {\bibinfo {author} {\bibfnamefont {T.}~\bibnamefont
  {Qiu}}, \bibinfo {author} {\bibfnamefont {J.}~\bibnamefont {Evslin}},
  \bibinfo {author} {\bibfnamefont {Y.-F.}\ \bibnamefont {Cai}}, \bibinfo
  {author} {\bibfnamefont {M.}~\bibnamefont {Li}},\ and\ \bibinfo {author}
  {\bibfnamefont {X.}~\bibnamefont {Zhang}},\ }\bibfield  {title} {\bibinfo
  {title} {{Bouncing Galileon Cosmologies}},\ }\href
  {https://doi.org/10.1088/1475-7516/2011/10/036} {\bibfield  {journal}
  {\bibinfo  {journal} {JCAP}\ }\textbf {\bibinfo {volume} {1110}},\ \bibinfo
  {pages} {036}},\ \Eprint {https://arxiv.org/abs/1108.0593} {arXiv:1108.0593
  [hep-th]} \BibitemShut {NoStop}%
\bibitem [{\citenamefont {Easson}\ \emph {et~al.}(2011)\citenamefont {Easson},
  \citenamefont {Sawicki},\ and\ \citenamefont {Vikman}}]{Easson:2011zy}%
  \BibitemOpen
  \bibfield  {author} {\bibinfo {author} {\bibfnamefont {D.~A.}\ \bibnamefont
  {Easson}}, \bibinfo {author} {\bibfnamefont {I.}~\bibnamefont {Sawicki}},\
  and\ \bibinfo {author} {\bibfnamefont {A.}~\bibnamefont {Vikman}},\
  }\bibfield  {title} {\bibinfo {title} {{G-Bounce}},\ }\href
  {https://doi.org/10.1088/1475-7516/2011/11/021} {\bibfield  {journal}
  {\bibinfo  {journal} {JCAP}\ }\textbf {\bibinfo {volume} {1111}},\ \bibinfo
  {pages} {021}},\ \Eprint {https://arxiv.org/abs/1109.1047} {arXiv:1109.1047
  [hep-th]} \BibitemShut {NoStop}%
\bibitem [{\citenamefont {Akama}\ and\ \citenamefont
  {Kobayashi}(2017)}]{Akama:2017jsa}%
  \BibitemOpen
  \bibfield  {author} {\bibinfo {author} {\bibfnamefont {S.}~\bibnamefont
  {Akama}}\ and\ \bibinfo {author} {\bibfnamefont {T.}~\bibnamefont
  {Kobayashi}},\ }\bibfield  {title} {\bibinfo {title} {{Generalized
  multi-Galileons, covariantized new terms, and the no-go theorem for
  nonsingular cosmologies}},\ }\href
  {https://doi.org/10.1103/PhysRevD.95.064011} {\bibfield  {journal} {\bibinfo
  {journal} {Phys. Rev.}\ }\textbf {\bibinfo {volume} {D95}},\ \bibinfo {pages}
  {064011} (\bibinfo {year} {2017})},\ \Eprint
  {https://arxiv.org/abs/1701.02926} {arXiv:1701.02926 [hep-th]} \BibitemShut
  {NoStop}%
\bibitem [{\citenamefont {Cai}\ \emph {et~al.}(2017)\citenamefont {Cai},
  \citenamefont {Li}, \citenamefont {Qiu},\ and\ \citenamefont
  {Piao}}]{Cai:2017tku}%
  \BibitemOpen
  \bibfield  {author} {\bibinfo {author} {\bibfnamefont {Y.}~\bibnamefont
  {Cai}}, \bibinfo {author} {\bibfnamefont {H.-G.}\ \bibnamefont {Li}},
  \bibinfo {author} {\bibfnamefont {T.}~\bibnamefont {Qiu}},\ and\ \bibinfo
  {author} {\bibfnamefont {Y.-S.}\ \bibnamefont {Piao}},\ }\bibfield  {title}
  {\bibinfo {title} {{The Effective Field Theory of nonsingular cosmology:
  II}},\ }\href {https://doi.org/10.1140/epjc/s10052-017-4938-y} {\bibfield
  {journal} {\bibinfo  {journal} {Eur. Phys. J.}\ }\textbf {\bibinfo {volume}
  {C77}},\ \bibinfo {pages} {369} (\bibinfo {year} {2017})},\ \Eprint
  {https://arxiv.org/abs/1701.04330} {arXiv:1701.04330 [gr-qc]} \BibitemShut
  {NoStop}%
\bibitem [{\citenamefont {Battarra}\ \emph {et~al.}(2014)\citenamefont
  {Battarra}, \citenamefont {Koehn}, \citenamefont {Lehners},\ and\
  \citenamefont {Ovrut}}]{Battarra:2014tga}%
  \BibitemOpen
  \bibfield  {author} {\bibinfo {author} {\bibfnamefont {L.}~\bibnamefont
  {Battarra}}, \bibinfo {author} {\bibfnamefont {M.}~\bibnamefont {Koehn}},
  \bibinfo {author} {\bibfnamefont {J.-L.}\ \bibnamefont {Lehners}},\ and\
  \bibinfo {author} {\bibfnamefont {B.~A.}\ \bibnamefont {Ovrut}},\ }\bibfield
  {title} {\bibinfo {title} {{Cosmological Perturbations Through a Non-Singular
  Ghost-Condensate/Galileon Bounce}},\ }\href
  {https://doi.org/10.1088/1475-7516/2014/07/007} {\bibfield  {journal}
  {\bibinfo  {journal} {JCAP}\ }\textbf {\bibinfo {volume} {1407}},\ \bibinfo
  {pages} {007}},\ \Eprint {https://arxiv.org/abs/1404.5067} {arXiv:1404.5067
  [hep-th]} \BibitemShut {NoStop}%
\bibitem [{\citenamefont {Ijjas}(2018)}]{Ijjas:2017pei}%
  \BibitemOpen
  \bibfield  {author} {\bibinfo {author} {\bibfnamefont {A.}~\bibnamefont
  {Ijjas}},\ }\bibfield  {title} {\bibinfo {title} {{Space-time slicing in
  Horndeski theories and its implications for non-singular bouncing
  solutions}},\ }\href {https://doi.org/10.1088/1475-7516/2018/02/007}
  {\bibfield  {journal} {\bibinfo  {journal} {JCAP}\ }\textbf {\bibinfo
  {volume} {1802}}\bibfield  {number} {\bibinfo  {number} { (02)},\ \bibinfo
  {pages} {007}},\ }\Eprint {https://arxiv.org/abs/1710.05990}
  {arXiv:1710.05990 [gr-qc]} \BibitemShut {NoStop}%
\bibitem [{\citenamefont {Ijjas}\ and\ \citenamefont
  {Steinhardt}(2016)}]{Ijjas:2016tpn}%
  \BibitemOpen
  \bibfield  {author} {\bibinfo {author} {\bibfnamefont {A.}~\bibnamefont
  {Ijjas}}\ and\ \bibinfo {author} {\bibfnamefont {P.~J.}\ \bibnamefont
  {Steinhardt}},\ }\bibfield  {title} {\bibinfo {title} {{Classically stable
  nonsingular cosmological bounces}},\ }\href
  {https://doi.org/10.1103/PhysRevLett.117.121304} {\bibfield  {journal}
  {\bibinfo  {journal} {Phys. Rev. Lett.}\ }\textbf {\bibinfo {volume} {117}},\
  \bibinfo {pages} {121304} (\bibinfo {year} {2016})},\ \Eprint
  {https://arxiv.org/abs/1606.08880} {arXiv:1606.08880 [gr-qc]} \BibitemShut
  {NoStop}%
\bibitem [{\citenamefont {Mironov}\ \emph {et~al.}(2018)\citenamefont
  {Mironov}, \citenamefont {Rubakov},\ and\ \citenamefont
  {Volkova}}]{Mironov:2018oec}%
  \BibitemOpen
  \bibfield  {author} {\bibinfo {author} {\bibfnamefont {S.}~\bibnamefont
  {Mironov}}, \bibinfo {author} {\bibfnamefont {V.}~\bibnamefont {Rubakov}},\
  and\ \bibinfo {author} {\bibfnamefont {V.}~\bibnamefont {Volkova}},\
  }\bibfield  {title} {\bibinfo {title} {{Bounce beyond Horndeski with GR
  asymptotics and $\gamma$-crossing}},\ }\href
  {https://doi.org/10.1088/1475-7516/2018/10/050} {\bibfield  {journal}
  {\bibinfo  {journal} {JCAP}\ }\textbf {\bibinfo {volume} {1810}}\bibfield
  {number} {\bibinfo  {number} { (10)},\ \bibinfo {pages} {050}},\ }\Eprint
  {https://arxiv.org/abs/1807.08361} {arXiv:1807.08361 [hep-th]} \BibitemShut
  {NoStop}%
\bibitem [{\citenamefont {Mukhanov}\ \emph {et~al.}(1992)\citenamefont
  {Mukhanov}, \citenamefont {Feldman},\ and\ \citenamefont
  {Brandenberger}}]{Mukhanov:1990me}%
  \BibitemOpen
  \bibfield  {author} {\bibinfo {author} {\bibfnamefont {V.~F.}\ \bibnamefont
  {Mukhanov}}, \bibinfo {author} {\bibfnamefont {H.~A.}\ \bibnamefont
  {Feldman}},\ and\ \bibinfo {author} {\bibfnamefont {R.~H.}\ \bibnamefont
  {Brandenberger}},\ }\bibfield  {title} {\bibinfo {title} {Theory of
  cosmological perturbations},\ }\href@noop {} {\bibfield  {journal} {\bibinfo
  {journal} {Physics Reports}\ }\textbf {\bibinfo {volume} {215}},\ \bibinfo
  {pages} {203} (\bibinfo {year} {1992})}\BibitemShut {NoStop}%
\bibitem [{\citenamefont {Sriramkumar}(2009)}]{Sriramkumar:2009kg}%
  \BibitemOpen
  \bibfield  {author} {\bibinfo {author} {\bibfnamefont {L.}~\bibnamefont
  {Sriramkumar}},\ }\bibfield  {title} {\bibinfo {title} {{An introduction to
  inflation and cosmological perturbation theory}},\ }\href@noop {} {\
  (\bibinfo {year} {2009})},\ \Eprint {https://arxiv.org/abs/0904.4584}
  {arXiv:0904.4584 [astro-ph.CO]} \BibitemShut {NoStop}%
\bibitem [{\citenamefont {Lehners}(2008)}]{Lehners:2008vx}%
  \BibitemOpen
  \bibfield  {author} {\bibinfo {author} {\bibfnamefont {J.-L.}\ \bibnamefont
  {Lehners}},\ }\bibfield  {title} {\bibinfo {title} {{Ekpyrotic and Cyclic
  Cosmology}},\ }\href {https://doi.org/10.1016/j.physrep.2008.06.001}
  {\bibfield  {journal} {\bibinfo  {journal} {Phys. Rept.}\ }\textbf {\bibinfo
  {volume} {465}},\ \bibinfo {pages} {223} (\bibinfo {year} {2008})},\ \Eprint
  {https://arxiv.org/abs/0806.1245} {arXiv:0806.1245 [astro-ph]} \BibitemShut
  {NoStop}%
\bibitem [{\citenamefont {Battefeld}\ and\ \citenamefont
  {Peter}(2015)}]{Battefeld:2014uga}%
  \BibitemOpen
  \bibfield  {author} {\bibinfo {author} {\bibfnamefont {D.}~\bibnamefont
  {Battefeld}}\ and\ \bibinfo {author} {\bibfnamefont {P.}~\bibnamefont
  {Peter}},\ }\bibfield  {title} {\bibinfo {title} {{A Critical Review of
  Classical Bouncing Cosmologies}},\ }\href
  {https://doi.org/10.1016/j.physrep.2014.12.004} {\bibfield  {journal}
  {\bibinfo  {journal} {Phys.Rept.}\ }\textbf {\bibinfo {volume} {571}},\
  \bibinfo {pages} {1} (\bibinfo {year} {2015})},\ \Eprint
  {https://arxiv.org/abs/1406.2790} {arXiv:1406.2790 [astro-ph.CO]}
  \BibitemShut {NoStop}%
\bibitem [{\citenamefont {Brandenberger}(2012)}]{Brandenberger:2012zb}%
  \BibitemOpen
  \bibfield  {author} {\bibinfo {author} {\bibfnamefont {R.~H.}\ \bibnamefont
  {Brandenberger}},\ }\bibfield  {title} {\bibinfo {title} {{The Matter Bounce
  Alternative to Inflationary Cosmology}},\ }\href@noop {} {\  (\bibinfo {year}
  {2012})},\ \Eprint {https://arxiv.org/abs/1206.4196} {arXiv:1206.4196
  [astro-ph.CO]} \BibitemShut {NoStop}%
\bibitem [{\citenamefont {Raveendran}\ \emph {et~al.}(2018)\citenamefont
  {Raveendran}, \citenamefont {Chowdhury},\ and\ \citenamefont
  {Sriramkumar}}]{Raveendran:2017vfx}%
  \BibitemOpen
  \bibfield  {author} {\bibinfo {author} {\bibfnamefont {R.~N.}\ \bibnamefont
  {Raveendran}}, \bibinfo {author} {\bibfnamefont {D.}~\bibnamefont
  {Chowdhury}},\ and\ \bibinfo {author} {\bibfnamefont {L.}~\bibnamefont
  {Sriramkumar}},\ }\bibfield  {title} {\bibinfo {title} {{Viable
  tensor-to-scalar ratio in a symmetric matter bounce}},\ }\href
  {https://doi.org/10.1088/1475-7516/2018/01/030} {\bibfield  {journal}
  {\bibinfo  {journal} {JCAP}\ }\textbf {\bibinfo {volume} {1801}}\bibfield
  {number} {\bibinfo  {number} { (01)},\ \bibinfo {pages} {030}},\ }\Eprint
  {https://arxiv.org/abs/1703.10061} {arXiv:1703.10061 [gr-qc]} \BibitemShut
  {NoStop}%
\end{thebibliography}%
\end{document}